\documentclass[10pt,aps,prb,twocolumn,showpacs,superscriptaddress,groupedaddress]{revtex4-1}
\usepackage[dvips]{graphics,graphicx}
\usepackage{bbold}
\usepackage{amsmath}
\usepackage[usenames, dvipsnames]{color}
\usepackage{epstopdf}
\usepackage{hyperref}
\usepackage[usenames]{color}
\usepackage{color}
\usepackage{colortbl}
\definecolor{linkcolor}{rgb}{0.8,0,0.2}
\definecolor{citecolor}{rgb}{0,0.6,0.2}
\definecolor{urlcolor}{rgb}{0,0,1}
\hypersetup{
colorlinks, linkcolor={linkcolor},
citecolor={citecolor}, urlcolor={urlcolor}}

\begin{document}

\title{Polarized bound state in the continuum 
\texorpdfstring{\\}{}
and resonances with tunable Q-factor in an anisotropic photonic crystal}
\author{Ivan V. Timofeev$^{1,2}$}
\email{tiv@iph.krasn.ru}
\author{Dmitrii N. Maksimov$^{1,3}$}
\author{Almas F. Sadreev$^1$}
\affiliation{
$^1$Kirensky Institute of Physics, Federal Research Center KSC SB
RAS, 660036, Krasnoyarsk, Russia\\
$^2$Siberian Federal University, 660041, Kransnoyarsk, Russia\\
$^3$Reshetnev Siberian State University of Science and Technology, 660037, Krasnoyarsk,
Russia}
\date{\today}
\begin{abstract}
{ We consider a one-dimensional photonic crystal composed of
alternating layers of isotropic and anisotropic dielectric
materials. Such a system has different band structures for
different polarizations of light. We demonstrate that if an
anisotropic defect layer is inserted into the structure, the
crystal can support an optical bound state in the continuum. By
tilting the principle dielectric axes of the defect layer relative
to those of the photonic crystal we observe a long-lived resonance
in the transmission spectrum. We derive an analytical expression
for the decay rate of the resonance that agrees well with the
numerical data by the Berreman anisotropic transfer matrix
approach. An experimental set-up with a liquid crystal defect
layer is proposed to tune the Q-factor of the resonance through
applying an external electric field. We speculate that the set-up
provides a simple and robust platform for observing optical bound
states in the continuum in the form of resonances with tunable
Q-factor.}

\end{abstract}
\maketitle

\section{Introduction}

The spectral properties of open systems are characterized by
resonant eigenvalues, which are typically complex numbers. It is
generally perceived that the imaginary part of a resonant
eigenvalue is due to the coupling of the resonant state to the
continuum of propagating eigenmodes corresponding to the
scattering channels. It might occurs, however, that under
variation of some parameters the imaginary part of a resonant
eigenvalue tends to zero, i.e. the resonant state becomes
decoupled from the scattering channels, whilst still embedded into
the spectrum of the extended states. Such continuum decoupled
states are source-free localized eigenmodes with infinite lifetime
known as bound states in the continuum (BICs) \cite{Review}.

The symmetry selection rules provide the simplest mechanism for cancelling the coupling of a bound
state to the continuum. The symmetry protected BICs in quantum waveguides were first proposed by Robnik in
a simple separable system with antisymmetric BICs embedded into the spectrum of symmetric propagating eigenstates
\cite{Robnik}. Later on the symmetry protected BICs were reported
in a cross-wire waveguide \cite{Schult} and a quantum dot subject to magnetic field \cite{Noeckel92}.

Nowadays we witness a surge of interest to BICs in field of
photonics, where BICs were observed in various set-ups with
periodical dielectric permittivity
\cite{Plotnik,Weimann13,Hsu13,Vicencio15,Sadrieva17,Xiao17}. In
particular, the studies on BICs are motivated by applications to
resonant light enhancement \cite{Mocella15, Yoon15, Bulgakov17}
and lasing \cite{Kodigala17,Bahari}. Another remarkable property
of BICs is the emergence of a collapsing Fano feature in its
parametric vicinity \cite{Kim,Shipman,Sadreev_BR2006,Blanchard16}, that can
be potentially employed to narrow-band filters
\cite{Foley14,Cui16}. To the best of our knowledge, optical
resonances with infinite lifetime in the $\Gamma$-point (i.e.
symmetry protected BICs) were first predicted in
\cite{Pacradouni00}. Another optical set-up supporting symmetry
protected BICs is a directional waveguide side-coupled with two
off-channel microcavities \cite{BS1} buried in the bulk of a
two-dimensional photonic crystal (PhC). That is the set-up
experimentally realized by Plotnik {\it et al} \cite{Plotnik}.
Recently, BICs have been studied in systems with dielectric
anisotropy \cite{Shipman2013,Torner} with the key idea to employ
the anisotropy for manipulating the frequency cut-offs  for
different polarizations of light in the ambient medium.

In this paper we propose a simple set-up for a symmetry protected BIC
localized in the vicinity of
an anisotropic defect layer (ADL) embedded into a one-dimensional anisotropic PhC.
The proposed set-up is similar to that from \cite{Shipman2013} with the only difference that the defect
layer in the above reference was taken isotropic. We will show that the addition of an ADL allows
us to provide an exact
analytical solution for a BIC and derive the decay rate
of the BIC related resonance in the transmission/reflection spectra.
An experimental set-up with a liquid crystal ADL
is proposed to control the Q-factor through applying an external electric field.

The paper is organized as follows: in Sec. \ref{Sec2} we review
the band structure of anisotropic PhCs, describe our model, and
numerically demonstrate the resonant feature associated with a
BIC. In Sec. \ref{Sec3} we present an exact analytical solution
for the BIC within the extraordinary waves stop-band. In Sec.
\ref{Sec4} we derive the decay rate of BIC-related resonance which
emerges in the scattering spectrum when the symmetry is broken by
application of ADL axes tilt. We confirm our findings with
numerical data in Sec. \ref{Sec5}. Finally, we conclude in Sec.
\ref{Sec6}.


\section{Model}\label{Sec2}
\begin{figure}
\includegraphics[scale=0.45]{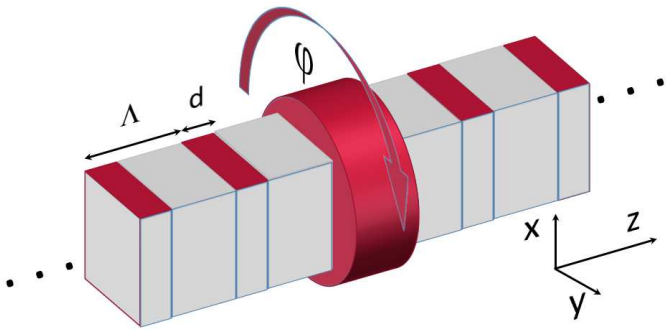}
\caption{One-dimensional PhC structure stacked of
alternating layers of an isotropic dielectric material with permittivity $\epsilon_o$ (gray)
and an anisotropic material with the permittivity components $\epsilon_o$ and $\epsilon_e$
(red). A defect layer with a tuneable permittivity tensor is inserted in the center of the structure.}
\label{fig1}
\end{figure}

We consider a one-dimensional PhC composed of alternating layers
of isotropic and anisotropic dielectric materials as shown in Fig. \ref{fig1}. The layers are stacked along the $z$-axis with
period $\Lambda$.
The isotropic layers are made of a dielectric material with permittivity $\epsilon_o$ and thickness
$\Lambda-d$, while the anisotropic layers have the principal dielectric axes aligned with
the $x$,$y$-axes with the corresponding permittivity components $\epsilon_e$, $\epsilon_o$. The thickness
of each anisotropic layer is $d$.

\begin{figure}[t]
\includegraphics[scale=1]{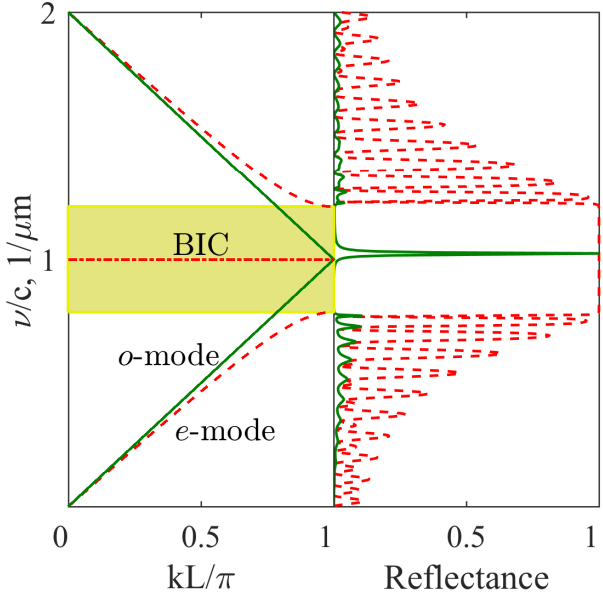}
\caption{ The band structure (left) and reflectance spectra
(right) of extraordinary (dash red) and ordinary waves (solid
green) in anisotropic PhC structure from Fig. \ref{fig1} with
$\nu$ as the linear frequency of the incident wave. The BIC
frequency is marked by a red dash-dot line. The yellow strip is
the photonic band gap. Dielectric permittivities of layers are
$\epsilon_e = 4;\epsilon_o = 1$. The widths $d = 0.125\, \mu m$,
$(\Lambda-d) = 0.250\, \mu m$ are optimized for normalized linear
frequency $\nu/c = 1/\lambda = 1\, \mu m^{-1}$ in the band gap
center. The defect layer has twice the thickness of the
anisotropic layers $2d=0.250\, \mu m$. The tilt angle
$\phi=\pi/8$.} \label{fig2}
\end{figure}

We start with Maxwell's equations for waves propagating along the $z$-axis. Assuming that the
wave vector is aligned with the $z$-axis the wave equations for the electric vector ${\bf E}=[E_x,E_y,0]$ are the following
\begin{align}\label{MEL}
\frac{\partial^2 E_{x}}{\partial z^2}+\frac{\omega^2}{c^2}\epsilon_oE_x=0, \notag \\
\frac{\partial^2 E_{y}}{\partial z^2}+\frac{\omega^2}{c^2}\epsilon(z)E_{y}=0,
\end{align}
where $\epsilon(z)$ is either $\epsilon_o$ within the isotropic layer, or $\epsilon_e$ within the
anisotropic ones. The solution for the ordinary $y$-polarized waves is obvious
\begin{equation}\label{ordinary}
E_y = E_0\exp{(i n_o k_0 z - i \omega t)},
\end{equation} where
$k_0 = \omega/c$ is the wave vector in vacuum, while $n_o=\sqrt{\epsilon_o}$ is the refractive index of
the isotropic material.
For the extraordinary $x$-polarized waves we have a piecewise continuus solution.
By using the transfer matrix approach the dispersion relation for the  extraordinary waves
is found as \cite{Markos,Yariv2007bk}
\begin{eqnarray*}
\cos{k_B \Lambda}=\cos{k_e d} \,\, \cos{k_0 (\Lambda-d)}- \\
\frac{1}{2} \left(
{\frac{n_e}{n_o}}+
{\frac{n_o}{n_e}}
\right)
\sin{k_e d} \,\, \sin{k_0 (\Lambda-d)},
\end{eqnarray*}
where $k_e = (\omega/c) \sqrt{\epsilon_e}$, $n_e=\sqrt{\epsilon_e}$, and $k_B$ is the Bloch vector.
The band structure is shown in Fig.\ref{fig2}(left).

According to Eq. (\ref{MEL}) the waves of different polarization do not mix in the PhC structure. The picture
becomes more involved if an anisotropic defect layer (ADL) of thickness $2d$ is inserted into the
center of the structure. In what follows we assume that the ADL is made of a material with the same principal
dielectric constants $\epsilon_o, \epsilon_e$, but the principal axes of the ADL are tilted with respect of the
principle axes of the bulk PhC by angle $\phi$ as shown in Fig. \ref{fig1}. We mention in passing that the set-up could
be implemented with a liquid crystal defect layer with the principal axes aligned with an external electric field.
The dielectric tensor of the ADL can be written as
\begin{equation}
\hat{\epsilon} = \left[\begin{array} {cc}
\epsilon_e \cos^2 \phi + \epsilon_o \sin^2 \phi & \sin 2\phi \;  (\epsilon_e-\epsilon_o)/2 \\
\sin 2\phi \; (\epsilon_e-\epsilon_o)/2 & \epsilon_e \sin^2 \phi + \epsilon_o \cos^2 \phi
\end{array} \right],
\label{eq:epsilon}
\end{equation}
where the tilt angle $\phi$ is the polar angle in the $x0y$-plane. For brevity
here and later on  we omit the $z$-components the
electromagnetic (EM) field, since $\hat{\epsilon}$ is a $2\times2$ matrix.

The whole system can be now viewed as a one-dimensional scattering
set-up with the ADL playing the role of the scattering center and
the PhC arms acting as semi-infinite waveguides. The resonant
properties of the system can be probed by an ordinary wave Eq.
(\ref{ordinary}) injected through the left arm. It is clear from
Eq. (\ref{eq:epsilon}) that if $\phi=0$ the system remains
transparent for the incident wave Eq. (\ref{ordinary}). If,
however, a tilt $\phi \neq 0$ of the dielectric axes is applied to
the ADL we expect a scattering solution with a mixture of both
polarizations in the vicinity of the ADL. The above speculation is
exemplified with numerical results in Fig. \ref{fig2} (right). The
numerical data are obtained with the Berreman transfer matrix
method \cite{Berreman1972} for the set of parameters collected in
the caption to Fig. \ref{fig2}. Most remarkably, one can see from
Fig. \ref{fig2} that the reflectance spectrum of the ordinary wave
exhibits a resonant feature in the middle of the extraordinary
waves stop-band. In what follows we will demonstrate that this
feature is induced by an $x$-polarized BIC which is converted to a
long-lived resonance (quasi-BIC) by the ADL axes tilt.
\section{Bound state in the continuum}\label{Sec3}

Let us  now construct a BIC solution for the ADL.
By definition the BIC is a source free solution localized in the vicinity of
the ADL. The solution must satisfy time-stationary Maxwell's equations
\begin{equation}\label{Maxwell}
\left\{\begin{array}{cc}
0 & \nabla \times \\
-\nabla \times & 0
\end{array}\right\}
\left\{\begin{array}{c}
{\bf E} \\
{\bf H}
\end{array} \right\}=
-ik_0
\left\{\begin{array}{c}
\hat{\epsilon} {\bf E} \\
{\bf H} \end{array}\right\},
\end{equation}
where $k_0=\omega/c$. Let us take the layers of equal
quarter-wavelength optical thickness for both materials
\begin{equation}\label{constraint}
d{n_e}=(\Lambda-d){n_o}=\lambda_0/4,
\end{equation}
with $\lambda_0=2\pi/k_0$ as the vacuum wavelength.
Notice that the above choice of parameters constraints
both the angular frequency of the BIC
\begin{equation}\label{BIC_frequency}
\omega=\frac{c\pi}{2dn_e},
\end{equation}
and the period of the PhC
\begin{equation}\label{BIC_period}
\Lambda=\frac{d(n_e+n_o)}{n_o}.
\end{equation}
In what follows we show that the specific choice of parameters
given by
 Eqs.(\ref{constraint},\ref{BIC_frequency},\ref{BIC_period}) allows us to construct
  a source-free solution that exponentially decays away from the ADL.
In the unperturbed case $\phi=0$ the BIC is polarized along the $x$-axis, hence for the EM field within the
ADL we can write
\begin{equation}\label{BIC}
\left.\
\begin{array}{l}
E_x^{(0)}(z)=\frac{1}{n_e}A\sin({n_e}k_0z), \\
H_y^{(0)}(z)=-iA\cos({n_e}k_0z).
\end{array}
\right.\
\end{equation}
The above solution obviously satisfies Eq. (\ref{Maxwell}) within the ADL. Let us
demonstrate that it can be extended into the PhC arms by matching the EM fields
on the boundaries between the anisotropic and isotropic layers. We denote the EM field components
in the isotropic layers adjacent to the ADL by $\bar{E}_x^{(0)}(z)$, and $\bar{H}_y^{(0)}(z)$. On the boundaries
of the ADL the following boundary conditions must be satisfied
\begin{equation}\label{BC}
\bar{E}_x^{(0)}(\pm d)={E}_x^{(0)}(\pm d), \
\bar{H}_y^{(0)}(\pm d)={H}_y^{(0)}(\pm d).
\end{equation}
Here we describe the wave matching for $z>0$ having in mind that
by construction the solution Eq. (\ref{BIC}) is antisymmetric with
respect to the $z$-axis. By using Eq. (\ref{BIC}) we can write
\begin{equation}\label{next}
\left.\
\begin{array}{l}
\bar{E}_x^{(0)}(z)=\frac{1}{n_e}A\sin[{n_o}k_0(\Lambda-z)], \\
\bar{H}_y^{(0)}(z)=iqA\cos[{n_o}k_0(\Lambda-z)],
\end{array}
\right.\
\end{equation}
where
$$q=\frac{n_o}{n_e}<1.$$
Re-iterating the matching procedure into the depth of the PhC arm
we find the EM fields within the $\mathrm{m_{th}}$ anisotropic and isotropic layers
\begin{equation}
\begin{array}{l}
E_x^{(m)}(z)=\frac{1}{n_e}(-1)^m q^mA\sin[{n_e}k_0(z-m\Lambda)], \\
H_y^{(m)}(z)=i(-1)^{m+1} q^mA\cos[{n_e}k_0(z-m\Lambda)],
\end{array}
\end{equation}
and
\begin{equation}
\begin{array}{l}
\bar{E}_x^{(m)}(z)=\frac{1}{n_e}(-1)^m q^mA\sin[{n_o}k_0((m+1)\Lambda-z)], \\
\bar{H}_y^{(m)}(z)=i(-1)^{m} q^{m+1}A\cos[{n_o}k_0((m+1)\Lambda-z)].
\end{array}
\end{equation}
One can see that we have found a localized solution which decays exponentially with $m$, Q.E.D.

Let us define the quantities  ${{\cal E}_m}$, $\bar{{\cal E}}_m$ as the energies stored in
the $\mathrm{m_{th}}, m=0,1,2,\ldots$ anisotropic and isotropic layers, correspondingly.
Notice that for $\phi=0$ one half of the ADL $m=0$ is identical to the next anisotropic layers in the PhC arm
$m=1,2,\ldots$.
The energies can be expressed through the integrals
\begin{equation}
{\cal E}_m=\frac{1}{8\pi}\int\limits_{m\Lambda}^{m\Lambda+d}dz \left[
\left.{\bf E}^{(m)}\right.^{\dagger}\hat{\epsilon}(z){\bf E}^{(m)}+
\left.{\bf H}^{(m)}\right.^{\dagger}{\bf H}^{(m)} \right],
\end{equation}
and
\begin{equation}
\bar{{\cal E}}_m=\frac{1}{8\pi}\int\limits_{m\Lambda+d}^{(m+1)\Lambda}dz \left[
\left.\bar{{\bf E}}^{(m)}\right.^{\dagger}\hat{\epsilon}(z)\bar{{\bf E}}^{(m)}+
\left.\bar{{\bf H}}^{(m)}\right.^{\dagger}\bar{{\bf H}}^{(m)} \right].
\end{equation}
The integrals can be evaluated as
\begin{equation}
{\cal E}_m=\frac{dA^2q^{2m}}{8\pi}, \ \bar{{\cal E}}_m=\frac{(\Lambda-d)A^2q^{2m+2}}{8\pi}=\frac{dA^2q^{2m+1}}{8\pi},
\end{equation}
where Eq. (\ref{constraint}) was used in the last step of the derivation.
The total energy stored in the BIC is, then, expressed through the following equation
\begin{equation}
{\cal E}=2\sum_{m=0}^{\infty}({\cal E}_m+\bar{{\cal E}}_m)=\frac{dA^2}{4\pi(1-q)}.
\end{equation}
By equating the total energy to unity we have
\begin{equation}\label{A}
A=\sqrt{\frac{4\pi(1-q)}{d}}.
\end{equation}
One can see that the BIC is now energy normalized in the whole space
for any value of the parameters $d, \epsilon_o, \epsilon_e$.
\begin{figure}
\includegraphics[scale=1]{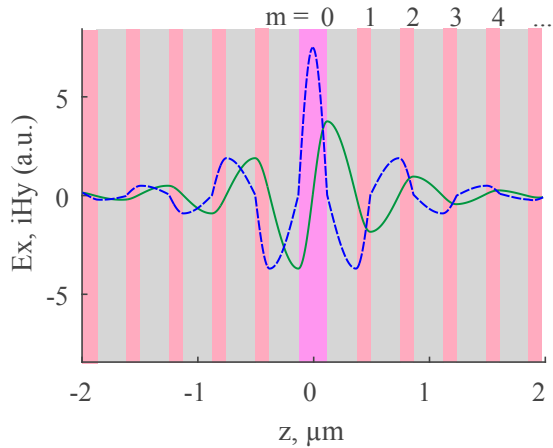}
\caption{BIC mode profile; The electric field $E_x $ -- solid
green, the magnetic field $H_y$ -- dash blue; tilt angle
$\phi=\pi/18$. The other parameters are the same as in
Fig.~\ref{fig2}.} \label{fig3}
\end{figure}

The BIC mode profile is plotted in Fig. \ref{fig3}. Notice, that
the structure is simmetric with respect to $x0z$- and $y0z$-plane
mirror reflections. The BIC polarization is orthogonal to the
polarization of propagating waves Eq. (\ref{ordinary}). This
orthogonality can be thought of as a case of symmetry protection.

\section{Decay rate}\label{Sec4}

As it was demonstrated in Sec. \ref{Sec2}, in case the principle
axes of the ADL are tilted by angle $\phi\neq 0$ the BIC becomes a
leaky mode (quasi-BIC) emerging in the reflection spectrum as a
sharp resonant feature. In this section we derive an analytical
expression for the energy decay rate $\Gamma$ which by definition
is the ratio of the power lost into the continuum of the ordinary
waves, $P$ to the energy stored in the BIC, $\mathcal{E}$
\begin{equation}
{\Gamma} = \frac{\mathcal{P}}{\mathcal{E}}.
\label{eq:tau_def}
\end{equation}
Taking into account that the BIC is energy normalized the only unknown quantity in Eq. (\ref{eq:tau_def}) is $\mathcal{P}$.
Next we will demonstrate how two different techniques could be used for finding $\mathcal{P}$.

\subsection{Wave matching}

We start with a heuristical  wave-matching approach, similar in spirit to that from \cite{Becchi04,Belyakov11,Timofeev2017_CMT}.
Let us consider the standing wave solution in the ADL inserted to the isotropic medium of adjacent layers. 
In the considered case the ADL produces no reflection for stationary incident waves of arbitrary polarization. 
The ordinary wave is not reflected because of matched impedances with equal dielectric permittivity. 
For the extraordinary wave the ADL is a half-wavelength layer according to
Eq.~(\ref{constraint}). 
The total reflection from both boundaries of a half-wavelength layer is zero, 
see Eqs.~(\ref{BIC}) and (\ref{next}) 
and Fig.~\ref{fig3} with the electric field maximum at the ADL boundary. 
So, the electric field amplitude inside the ADL and in the adjacent layers is the same.
Thus, the leakage is due to transmission rather than reflection.

The standing wave can be decomposed into incident and
outgoing waves with respect to the ADL with electric field amplitudes $A_{(\mp)}$
$$A_{(\mp)}=\frac{1}{n_e}\sqrt{\frac{\pi(1-q)}{d}}.$$
One can write the matching condition
with ordinary and extraordinary waves within the tilted ADL
\begin{equation}\label{matching}
\begin{array}{l}
E_e(-d) = A_{(+)} \cos \phi \\
E_o(-d) = A_{(+)} \sin \phi,
\end{array}
\end{equation}
where $E_e$, and $E_o$ are the amplitudes of ordinary and extraordinary waves propagating to the right within the ADL
at the left interface $z=-d$.
After propagating to the right interface $z=d$ those waves accumulate phase
\begin{equation}\label{phase}
E_{e,o}(d) = E_{e,o}(-d)  \exp(2id{n_{e,o}} k_0).
\end{equation}
Projecting back onto the $y$-axis and combining the above
equations (\ref{matching}) and (\ref{phase}) we have for the amplitude
of the outgoing wave in the right PhC arm
\begin{equation}
|E_y| = \frac{1}{n_e}\sqrt{\frac{\pi(1-q)}{d}} \sin (2\phi) \sin \left(dk_0 ({n_e}-{n_o})\right).
\label{eq:E_y}
\end{equation}
Then, the Poynting vector amplitude of the outgoing wave can be
written as
\begin{equation}
P = c n_o |E_y^2|/4 \pi.
\label{eq:Power}
\end{equation}
The total energy loss must be obviously doubled to account for leakage into the
left PhC arm. Hence, by using Eqs. (\ref{BIC_frequency},\ref{eq:E_y}) and (\ref{eq:Power}) we have for
the power lost in unit of time
\begin{equation}\label{final_wave}
\mathcal{P}=\frac{c\sin(2\phi)^2q^2(1-q)}{2d{n_o}}\cos^2\left(\frac{\pi q}{2}\right).
\end{equation}
At this point we would like to emphasise the obvious violation of a half-wavelength condition Eq.~(\ref{constraint}) by the ADL tilt. 
When the tilt is small the quasi-BIC frequency shift is proportional to $\phi^2$, 
and the residual reflection amplitude has the same order of smallness. 
Then the reflection amplitude projection onto the $y$-axis is proportional to $\phi^3$. 
Thus, the expression Eq. (\ref{final_wave}) is accurate up to $\mathcal{O}(\phi^4) \sim \phi^3 sin(2\phi)$.

\subsection{Time-dependant perturbation theory}

To obtain an analytical expression for $\mathcal{P}$ in a more rigorous manner we will apply the time-dependant perturbation theory
assuming that initially the EM energy is loaded into the BIC with $\phi=0$. Then, at the moment $t=0$
an axes tilt is applied to the system, and the energy stored in the BIC leaks into the continuum of the
ordinary waves. Expanding the dielectric tensor in the powers of $\phi$ one finds
\begin{equation}\label{expansion}
\hat{\epsilon}(z,t)=\hat{\epsilon}_0(z)+\theta(t)\theta(d-|z|)\hat{\epsilon}_1+\mathcal{O}(\phi^2),
\end{equation}
where $\hat{\epsilon}_0(z)$ is the dielectric tensor of the unperturbed system, $\theta(\ldots)$ -
the Heaviside step function, and $\hat{\epsilon}_1$ can be found from Eq. (\ref{eq:epsilon}) as
\begin{equation}\label{tensor}
\hat{\epsilon}_1=-
{\phi}{\epsilon_e}\left\{
\begin{array}{cc}
0 & 1-q^2 \\
1-q^2 & 0
\end{array}
\right\}.
\end{equation}
Using the notation
\begin{equation}
\hat{H}=\left\{
\begin{array}{cc}
0 & \nabla \times \\
-\nabla \times & 0
\end{array}
\right\}
\end{equation}
we can write temporal Maxwell's equations
\begin{equation}{\label{temporal_solution}}
\hat{H}{\bf \Psi}(t)=\frac{\partial}{\partial t}
\left\{[\hat{\epsilon}_0(z)+\theta(t)\theta(d-|z|)\hat{\epsilon}_1]{\bf \Psi}(t)\right\},
\end{equation}
where ${\bf \Psi}$ is $4\times1$ vector ${\bf \Psi}=[E_x, E_y, H_x, H_y]$.
To proceed we write the solutions of the continuous spectrum as
\begin{equation}
{\bf \Psi}(k,z)= [0, E_y(k,z), H_x(k,z), 0] \\
\end{equation}
with
\begin{equation}
\begin{array}{l}
E_y(k,z)=\sqrt{\frac{2}{ n_o}} e^{in_okz}, \\
H_x(k,z)=\sqrt{2n_o} e^{in_okz}.
\end{array}
\end{equation}
The solutions are normalized to the Dirac delta
$$\frac{1}{8\pi}\int\limits_{-\infty}^{\infty}{\bf \Psi}(k,z)^{\dagger} \hat{\epsilon}_0(z){\bf \Psi}(k',z)dz=\delta(k-k').$$

Now we have all necessary ingredients for the time-dependant perturbation theory \cite{Landau58a} at our disposal.
According to the first order time-dependant perturbation theory the solution is given by
\begin{equation}\label{psolution}
{\bf \Psi}(t,z)={\bf \Psi}_0(z)e^{-ick_0t}+\int\limits_{-\infty}^{\infty} b(k,t){\bf \Psi}(k,z)e^{-ic|k|t}+\mathcal{O}(\phi^2),
\end{equation}
where $b(k,t)\sim \phi$, and ${\bf \Psi}_0(z)$ is the BIC mode profile. After substituting the above into Eq. (\ref{temporal_solution})
we find
\begin{equation}\label{amplitude}
\frac{\partial b(k,t)}{\partial t}=[\delta(t)-ick_0\theta(t)]e^{ic(|k|-k_0)t}V(k)
\end{equation}
with
\begin{equation}\label{V_n}
V(k)=\int\limits_{-d}^{d}dz{\bf \Psi}^{\dagger}(k,z)\hat{\epsilon}_1{\bf \Psi}_{(0)}(z).
\end{equation}
Evaluating the above integral we have
\begin{equation}
V(k)=-(1-q^2)n_e\phi\sqrt{\frac{1-q}{8\pi n_0d}}f(k,k_0),
\end{equation}
where
\begin{equation}
f(k,k_0)=\frac{\sin[d({n_o}k-{n_e}k_0)]}{{n_o}k-{n_e}k_0}-
\frac{\sin[d({n_o}k+{n_e}k_0)]}{{n_o}k+{n_e}k_0}.
\end{equation}
The energy lost to the continuum per unit of time can be now found as
\begin{equation}\label{loss}
\mathcal{P}(t)=\frac{1}{8\pi}\frac{\partial}{\partial t}\int
\limits_{-\infty}^{\infty}dz\left[{\bf \Psi}^{\dagger}(t)\hat{\epsilon}_0{\bf \Psi}(t)\right].
\end{equation}
Notice that the $\mathcal{P}(t)$ is time-dependant. Hence, we define time-averaged energy loss in the following
manner
$$\mathcal{P}=\lim_{T \rightarrow \infty}\frac{1}{T}\int\limits_0^{T}dt\mathcal{P}(t).$$
The above definition may appear superfluous, however, by subsisting into it Eqs. (\ref{loss}) and (\ref{temporal_solution}) and
applying the normalization condition we find a useful expression that will eventually simplify the further
analysis
\begin{equation}\label{power}
\mathcal{P}=\lim_{T \rightarrow \infty}\frac{1}{T}\int\limits_0^{T}dt\int\limits_{-\infty}^{\infty}
dk\left(b^{*}(k)\frac{\partial b(k)}{\partial t}
+\frac{\partial b^{*}(k)}{\partial t} b(k)\right)+\mathcal{O}(\phi^4),
\end{equation}
where the terms $\mathcal{O}(\phi^3)$ are dropped off since $\mathcal{P}$ is symmetric with respect to the sign of $\phi$.
Substituting Eq. (\ref{amplitude}) into Eq. (\ref{power}) and using L'Hopital's rule one obtains
\begin{align}
\mathcal{P}=[(1-q^2)ck_0\phi]^2{\frac{n_e^2(1-q)}{4\pi n_od}}\times\notag \\
\lim_{T \rightarrow \infty}\int\limits_0^T
dt\int\limits_{-\infty}^{\infty}dk \cos[tc(k_0-|k|)]f(k,k_0)^2+\mathcal{O}(\phi^4).
\end{align}
By recollecting  the identity
$$\delta(x-x')=\frac{1}{\pi c}\int\limits_0^{\infty}dt\cos[tc(x-x')]$$
we obtain
\begin{equation}
\mathcal{P}=[(1-q^2)k_0\phi]^2\frac{cn_e^2(1-q)}{2n_od}
f(k_0,k_0)^2+\mathcal{O}(\phi^4).
\end{equation}
Next by using Eq. (\ref{BIC_frequency}) we have
\begin{equation}
\begin{array}{l}
f(k_0,k_0)^2=\frac{1}{k_0^2n_e^2}\left\{\frac{\sin\left[(1-q)\frac{\pi}{2} \right]}{1-q}
-\frac{\sin\left[(1+q)\frac{\pi}{2} \right]}{1+q} \right\}^2\\
=\frac{4q^2}{k_0^2n_e^2(1-q^2)^2}\cos^2\left(\frac{\pi q}{2}\right)
\end{array}
\end{equation}
and finally
\begin{equation}\label{final}
\mathcal{P}=\frac{2c\phi^2q^2(1-q)}{dn_o}\cos^2\left(\frac{\pi q}{2}\right)+\mathcal{O}(\phi^4).
\end{equation}
One can see that Eqs. (\ref{final_wave}) and (\ref{final}) differ by $\mathcal{O}(\phi^4)$.

\section{Numerical results}\label{Sec5}
\begin{figure}[t]
\includegraphics[width=0.45\textwidth]{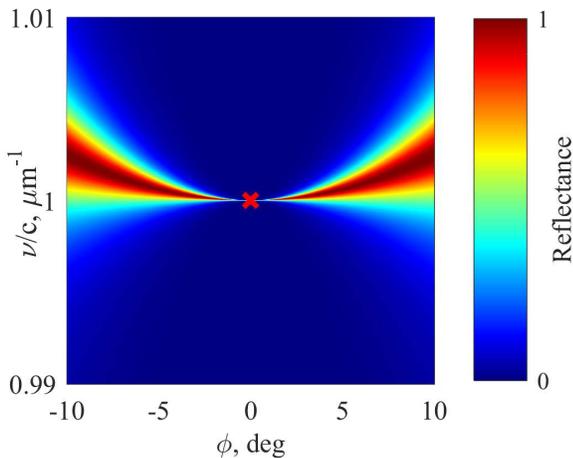}
\caption{%
Reflectance $|\rho|^2$ versus tilt angle $\phi$ and linear
frequency $\nu$. The other parameters are the same as in Fig.
\ref{fig1}. The red cross in the center corresponds to the BIC.}
\label{fig4}
\end{figure}
In this section we establish a link between the decay rate
$\Gamma$, Eq. (\ref{eq:tau_def}) and the numerical data on the
transmission and reflection spectra.  The scattering problem under
consideration poses a typical case of two pathways transmission
problem, where the direct path is identified with the incident
ordinary wave penetrating through the ADL from the left to the
right PhC arm. The second pathway is through the resonant
excitation of the quasi-BIC. Such scattering problem was
thoroughly analyzed by Suh, Wang, and Fan \cite{Suh04} in the
framework of the Coupled Mode Theory \cite{Manolatou1999}. The
general expression for the reflection/transmission amplitudes was
obtained as
\begin{equation}\label{transmission}
\rho = \frac{i\gamma}{(\omega_r-\omega) +i\gamma}, \ \tau=1-\rho,
\end{equation}
where $\omega_r$ is the position of the resonance, and $\gamma$ is
the imaginary part of the resonant frequency. Remarkably, the
transmission amplitude $\tau$ exhibits a transmission zero at
$\omega=\omega_r$ which could be understood as a consequence of a
full destructive interference between the two transmission
pathways. On the other hand, the reflection amplitude $\rho$ is
simply a Lorentzian of the width $2\gamma$. Taking into account
that the energy relaxation time of a resonant state is given by
$${\Gamma}={2\gamma},$$
we find a link between Eqs. (\ref{eq:tau_def},\ref{final_wave}) and (\ref{final}), and the resonant width in the
reflection spectrum as
\begin{equation}
\Delta\nu=\frac{\gamma}{\pi},
\end{equation}
where $\nu$ is the linear frequency.

\begin{figure}[t]
\includegraphics[width=0.45\textwidth]{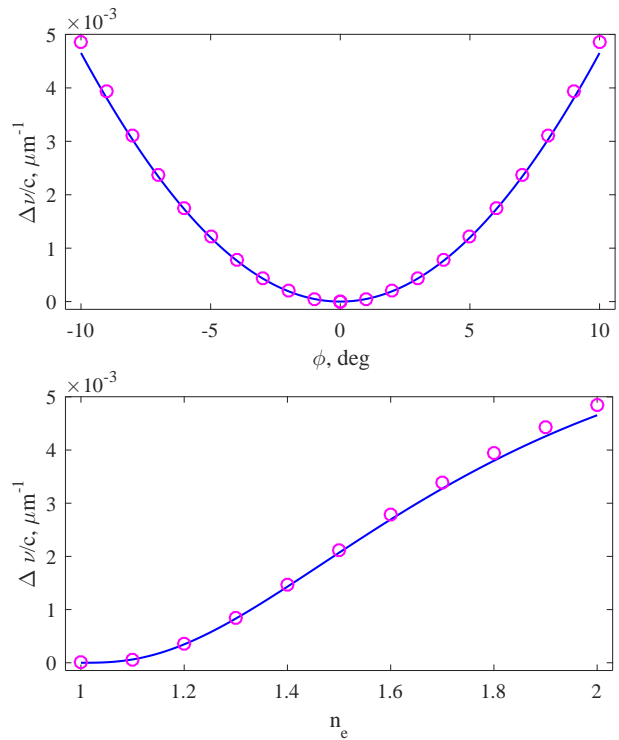}
\caption{%
Resonance width $\Delta\nu$ vs. tilt angle $\phi$ for $n_e=2$ (top), and the refraction index $n_e$ for $\phi=\pi/18$
(bottom). The geometry of the system is the same as in Fig. \ref{fig1}.
The numerical data are shown by red circles, the analytical results obtained from Eq.~(\ref{final_wave}) are shown by blue solid line.}
\label{fig5}
\end{figure}
In Fig. \ref{fig4} we present the reflectance vs. the tilt angle
$\phi$ and linear frequency of the incident wave $\nu$. In the BIC
point we observe a collapse of the resonance as its width turns to
zero. The numerical data from Fig. \ref{fig4} were used for
comparing the resonance widths with our analytical predictions.
The results are shown in Fig. \ref{fig5}, where one can see a good
agreement between theory and numerical experiment.

\section{Conclusion}\label{Sec6}

Most of theoretical works on PhCs rely on various numerical
techniques \cite{BS1,Torner,Venakides03,Marinica08,Gao16} for
finding BIC frequencies and mode profiles. So far, to the best of
our knowledge, the only model with analytical solution for optical
BICs was photonic Lieb lattice \cite{Vicencio15, MurPetit2014}. Here we have
found an exact analytical solution for an optical BIC in an
anisotropic defect layer embedded into an anisotropic PhC.
Moreover, the decay rate of a qausi-BIC resonance in the system
with tilted principal axes of the defect layer was computed with
the use of the time-dependant perturbation theory and the
transmission/reflection spectra are explained through the coupled
mode approach. A simple experimental set-up with a liquid crystal
defect layer is proposed to tune the Q-factor through applying an
external low-frequency electric field. The question of BICs
tunability has been previously discussed in the literature
\cite{Ni16} with the optical properties of the system changing
under variation of the thickness of dielectric slabs. This
approach, however, would require re-fabrication of the BIC
supporting structure. The idea of using the optoelectronic effect
for manipulating the Q-factor has already been applied to liquid
metacrystals \cite{Piccardi14, Zharov14} and microring resonantors
\cite{Wang07, Guarino07, Wang14}. We speculate that the analytical
approaches proposed in this work may find application to various
set-ups with optoelectronically controlled Q-factors.

\begin{acknowledgments}
This work was partially supported through RFBR grants
N17-52-45072, and N17-42-240464. We appreciate discussions with E.N.
Bulgakov and K.N. Pichugin.
\end{acknowledgments}

\bibliography{Liquid_Crystal}

\end{document}